\documentclass[a4paper]{article}

\usepackage{INTERSPEECH2019}
\usepackage{multirow}
\usepackage{hyperref}       
\usepackage{url}            

\title{A Fully Time-domain Neural Model for Subband-based Speech Synthesizer}
\name{Azam ~Rabiee$^1$, Geonmin ~Kim$^2$, Tae-Ho ~Kim$^1$, Soo-Young ~Lee$^1$}
\address{
  $^1$KAIST Institute for Artificial Intelligence \quad $^2$Department of Electrical Engineering\\
  Korea Advanced Institute of Science and Technology, Daejeon, Korea 
}
\email{azrabiee@kaist.ac.kr, gmkim90@gmail.com, ktho22@kaist.ac.kr, sy-lee@kaist.ac.kr}

\begin{document}

\maketitle
\begin{abstract}
  This paper introduces a deep neural network model for subband-based speech synthesizer. The model benefits from the short bandwidth of the subband signals to reduce the complexity of the time-domain speech generator. We employed the multi-level wavelet analysis/synthesis to decompose/reconstruct the signal into subbands in time domain. Inspired from the WaveNet, a convolutional neural network (CNN) model predicts subband speech signals fully in time domain. Due to the short bandwidth of the subbands, a simple network architecture is enough to train the simple patterns of the subbands accurately. In the ground truth experiments with teacher-forcing, the subband synthesizer outperforms the fullband model significantly in terms of both subjective and objective measures. In addition, by conditioning the model on the phoneme sequence using a pronunciation dictionary, we have achieved the fully time-domain neural model for subband-based text-to-speech (TTS) synthesizer, which is nearly end-to-end. The generated speech of the subband TTS shows comparable quality as the fullband one with a slighter network architecture for each subband. 
\end{abstract}


\section{Introduction}

Text-to-speech (TTS) techniques have been started from concatenative synthesis [1], [2] to statistical parametric speech synthesis [3]–[5], and eventually end-to-end fully neural network based models [6], [7]. Recent speech synthesizers have employed giant neural networks and high configuration GPUs to achieve remarkable success in more natural and fast speech generation. Of such models, WaveNet [8] has achieved the most natural generated speech that significantly closes the gap with human. As a deep generative network, WaveNet directly models the raw audio waveform, which has changed the existing paradigms. It made a paradigm to absorb a tremendous amount of attention for sequential modeling [9], speech enhancement [10], [11], and vocoder, which is the wave synthesizer from acoustic features [12]–[16]. 

Thanks to its convolutional structure, WaveNet benefits from parallel computing in train. However, the generation is still a sequential sample-by-sample process. Thus, due to the very high temporal resolution of speech signals (at least 16000 samples per second), the vanilla WaveNet suffers from the long generation time. Therefore, fast [17], parallel [18], [19], and glow-based [20] models are introduced. The fast model is an efficient implementation that removes redundant convolutional operations by caching them. While parallel and glow-based models utilized distillation and normalizing flow, respectively, which lead to the speech synthesis faster than real-time.

Unlike the huge network required in the parallel and glow-based models, some studies benefit from subband decomposition to reduce the complexity. Previously, a hybrid TTS [21] applied HMM-based and waveform-based synthesis for low and high frequencies, respectively. However, the TTS suffers from the drawbacks of the HMM-based models and the overall performance is not satisfying. In addition, a subband WaveNet vocoder [22] is presented using a frequency filterbank analysis. However, to have a TTS based on the subband vocoder, separate acoustic and linguistic models are required.

Similar to [22], the aim of this research is to break down the WaveNet architecture into smaller networks for each subband of the speech signal. The benefits of the subband model is the reduced computational complexity and the feasibility of training accurately for each subband due to their short bandwidth. Unlike the frequency domain filterbank used in [21] and [22], wavelet decomposition is used in this study. In addition, the similar morphological structure of the dilated convolutions in WaveNet and the wavelet transform has inspired us to use the wavelet. Thus, the innovation is utilizing the wavelet analysis to decompose the time-domain speech signal $s(t)$ into subbands $s_l (t) (l=1,…,L)$. Then, an integrated model generates each subband signal in parallel. The subband signal generator is based on the fast WaveNet [17].  

Even though many recent studies utilized the WaveNet as a vocoder, we believe that converting the spectrogram information to waveform is an inverse spectrogram process and may not necessarily need such a huge architecture. Instead, our hypothesis is that the WaveNet is able to perform some parts of the TTS front stage, as well. In addition, a single integrated model is likely to be more stable empirically and less-sensitive for parameter tuning than a multi-stage model, in which separate parameter tuning is required for each stage and the errors may add up [6], [23]. Hence, another contribution of this paper is that by simply conditioning the proposed model on the phoneme sequence and benefitting from an encoder, we have achieved the fully time-domain neural model for subband-based TTS. Section \ref{proposed} describes the proposed subband speech synthesizer. Section \ref{experiments}  explains our experiments and results. Finally, conclusion comes in Section \ref{conclusion}.     

\section{Proposed subband speech synthesizer}
\label{proposed}

The aim of this paper is to reduce the complexity of the time-domain TTS by decomposing the fullband speech signal $s$ into the subbands $s_l$ ($l=1,..., L$) using the wavelet analysis (detailed in Subsection \ref{proposed-dec-rec}). Due to the short bandwidth of the subbands, the structure of each subband generator is much slighter than the fullband one. Moreover, our hypothesis is that subband generators are more accurate as they are trained for the localized frequency patterns. Furthermore, our designed model benefits from the parallel processing in subband signal generators. When the subband signals are generated according to the corresponding conditional features using the localized TTS, then the wavelet synthesis reconstructs the fullband signal.  

In the designed model depicted in Figure 1, given the conditional features $c$, an encoder extracts the latent variables $h$ for generating samples conditioning on them (detailed in Subsection \ref{proposed-cond}). If the conditional features are linguistic features such as character or phoneme sequence, then the latent features would be linguistic features to make the model as a TTS.

The main part of the model is the autoregressive signal generators, shown by the outer dashed blocks in Figure 1. Each generator is in charge of modeling the probability distribution of each subband similar to the WaveNet (detailed in Subsection \ref{proposed-subband-gen}).

 In the training phase, the loss is defined by summation of the subband losses, which is the cross-entropy of the estimated and target subband signal, as 
\begin{equation}
loss=-\sum_{l=1}^{L}\mathbb{E}_{p_l}[\log{q_l}],
\end{equation}

in which $p_l$ and $q_l$ are the probability distribution of $s_l (t)$ and $\hat{s}_l (t)$, respectively. Since it is a probabilistic model, the generation model estimates the $t^{th}$ sample of each subband by sampling the learned probability distribution. 

With such a loss, the shared module, i.e. encoder, can train efficiently according to all the frequency bins. In addition, it is not timely efficient to train a separate network per subband, and then combine them. The integrated model in Figure 1 benefits from the parallel computation, along with any possible future subband integration via more shared modules for additional attributes such as speaker identity, prosody, speaking style, recording channel and noise levels.
\begin{figure}
  \centering
    \includegraphics[width=0.45\textwidth]{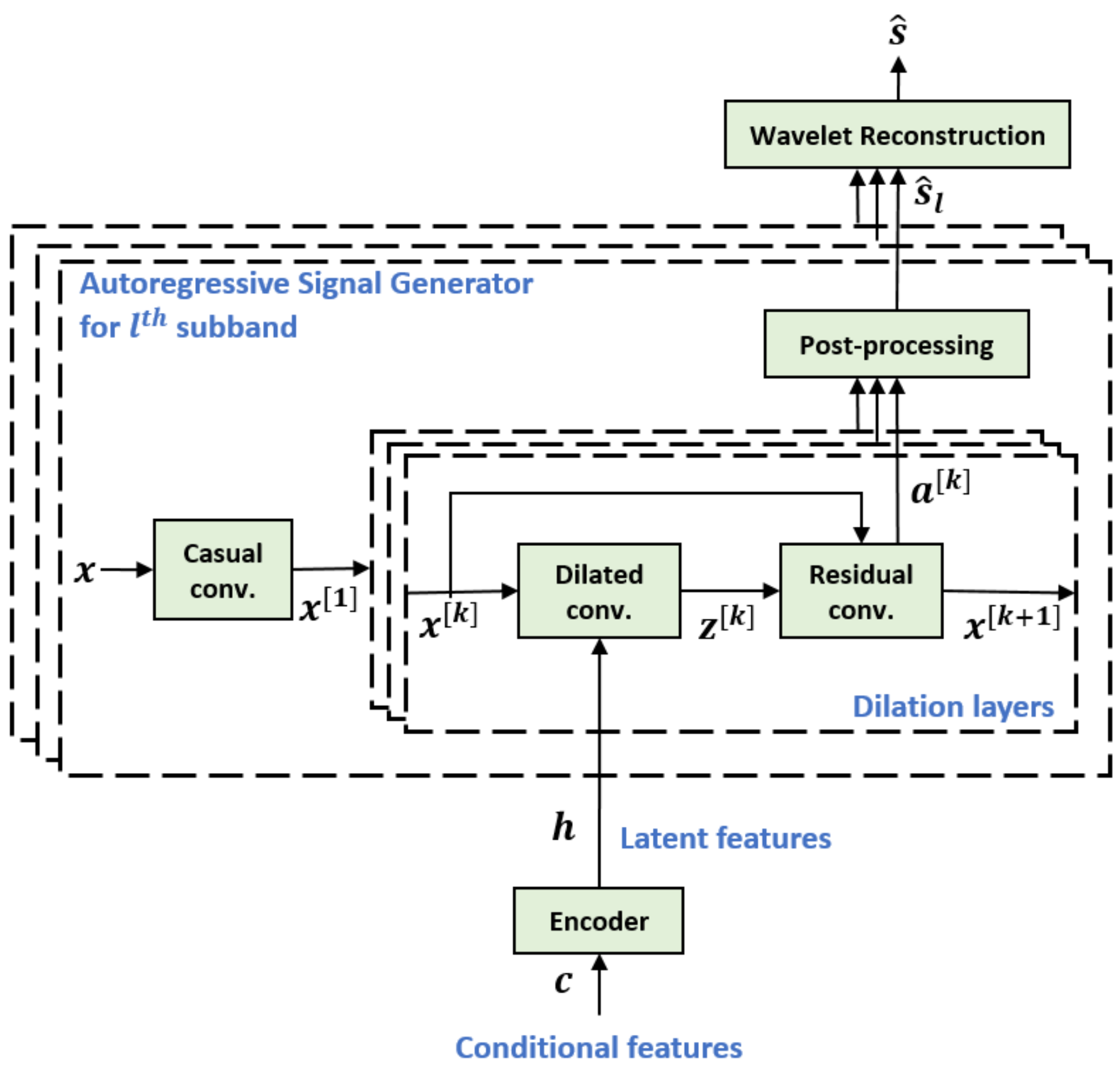}
    \caption{Schematic diagram of the proposed time-domain subband-based speech synthesizer. The model is trained to estimate subband signals $s_l$ conditioning on the latent variable $h$ extracted from conditional features $c$ and the previous time samples of the subband signal $x$. Linguistic and acoustic features can feed to the model as the conditional features for TTS.}
\end{figure}

\subsection{Subband decomposition/reconstruction}
\label{proposed-dec-rec}
A set of analysis filters can decompose speech signal $s(t)$ into subbands $s_l (t)$, and their paired synthesis filters are able to reconstruct back the fullband signal. The proposed synthesizer utilizes multi-level orthogonal time-domain wavelet as follows,

\begin{equation}
 \begin{cases}
  u_l (t)=u_{l-1} (t)*\varphi_l (t) \\
  s_l (t)=u_{l-1} (t)*\psi_l (t)
 \end{cases}
\end{equation}

where $\varphi_l(t)$ and $\psi_l (t)$ are Daubechies scaling (low-pass) and mother wavelet (high-pass) functions [24], respectively. Moreover, $l=1,...,L$ refers to the wavelet level and $u_0 (t)=s(t)$. The downsampling is omitted in every level of the wavelet transform because the downsampling widens the bandwidth, which needs more complex network to train. In addition, it decreases the size of the dataset. As there is no data like more data for the training, we ignored the downsampling after each layer.  

Reasons for selecting the wavelet transform rather than the short time Fourier transform (STFT) filterbank are as follows. First, the wavelet transform is very robust for reconstruction [25]. Corruption of the wavelet coefficients will only affect the reconstructed signal locally near the perturbed position, while the STFT will spread out the error everywhere in time. Second, output of the Fourier analysis filters are complex. Most of the spectrogram-based speech synthesizers ignore modeling the phase spectrogram [6], while the Fourier synthesis filters are sensitive to phase errors. Therefore, compared to the wavelet, the STFT models are unable to reconstruct the phase correctly. Third, the logarithmic spectral resolution of the wavelet are more compatible with the nature of speech compared to the uniform tiling of the spectrogram. Due to the nonlinear bandwidth divisions of the wavelet, high frequencies (e.g. above 4 kHz for 16 kHz sampling rate) fall in one subband. Whereas, there are fine divisions for the low frequencies. Later in the experiments, we will see the signal-to-noise ratio (SNR) of the consecutive decomposition and reconstruction is about 41 dB, in which the noise is hardly sensible by the human ear.

\subsection{Conditional/latent features}
\label{proposed-cond}
A variety of conditional features can be fed to the model for different possible applications such as vocoder or TTS. Of such features, we use phoneme sequence produced by a text normalization and lexicon to have a TTS model. The text normalization contains 1) converting to the lower-case characters, 2) removing the special symbols, and 3) converting numbers to text. The phoneme sequence speeds up the training [26]. As shown in Figure 1 by the encoder block, a number of convolutional layers along time axis can extract the linguistic features implicitly. The activation of the last layer, denoted by latent features $h$, is used for the generators. In fact, the encoder plays the role of the linguistic model for TTS, which is significant as shown in Figure 2. We have examined the fullband model without the encoder, which is in fact the original WaveNet conditioning on phoneme sequence; but the results were worse as the features were not enough for the training. Comparing fullband synthesis with and without the encoder module shows the importance of the module for automatic linguistic feature extraction in the proposed model. 

\begin{figure}
  \centering
    \includegraphics[width=0.45\textwidth]{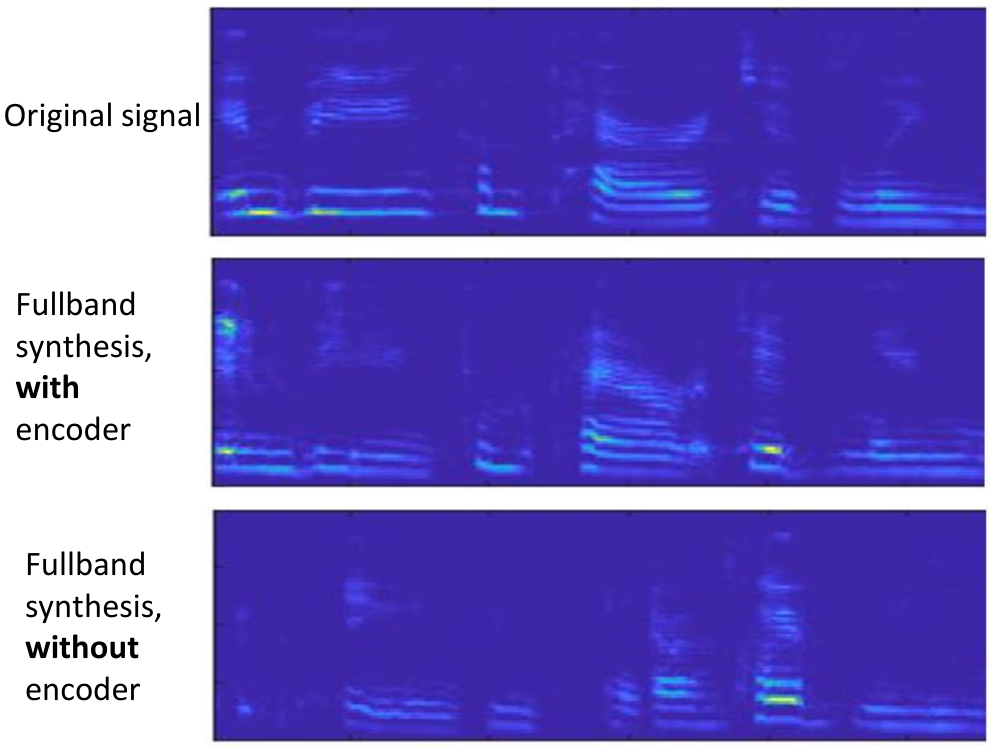}
    \caption{The Mel spectrograms show that the role of the encoder ($h=enc(c)$) is significant. The sample is selected randomly in the test set.}
\end{figure}

\subsection{Subband autoregressive signal generator}
\label{proposed-subband-gen}

The subband generator has a similar architecture as the WaveNet. Unlike the WaveNet, our autoregressive signal generator is in charge of generating subband signals. The model estimates the posterior probability of each subband time-sample $x_t$ conditioned on the previous samples, $x_{<t}$ and some latent features $h_t$ as $p(x_t |x_{<t},h_t)$.	

As shown in Figure 1, stacks of $K$ dilation layers in the dilations area perform dilated convolutions, residual connections, and skip connections. Note that the superscripts in Figure 1 show the layer index $(k=1,...,K)$. Convolutions with holes, as the dilated convolution layers, process the input in a fine to coarse scale with fewer weight parameters in the sufficient receptive field size. Thus, the output $a^{[k]}$ contains various latent feature of the input in different scales. The morphological structure of the dilated convolutions resemble the wavelet transform. In fact, with a specific set of weights, the first dilation layer can resemble the first level of the wavelet transform. Hence, the first layer mostly models the high frequency features, likewise, the higher dilations for the low frequencies. Therefore, a stack of $r$ repeats of $1,2,4,…,2^n$ dilations for modeling the fullband signal could be equivalent to $r$ repeats of one dilation layer for each wavelet subband. Therefore, in our experiments with subband signals, the number of dilation layers $K$ is much smaller than the original fullband WaveNet.

\section{Experiments}
\label{experiments}

We used the TTS benchmark dataset LJ Speech\footnote{https://keithito.com/LJ-Speech-Dataset/} consisting of 13,100 short audio clips uttered by a female speaker, varying in length from 1 to 10 seconds, recorded in 16kHz sampling rate. We kept around 11 minutes of the speech signals (100 utterances) for test, which was not included in the train. The training set lengths more than 23 hours after the silence removal using voice activity detector (VAD). 

\subsection{Parameter settings}
\label{exp-param}

The subband decomposition is performed by Daubechies wavelet \verb+db10+ for eight levels (L=8). Subband amplitude normalization is unavoidable because of the quantization in generator.

We found the Carnegie Mellon university pronouncing dictionary (CMUdict)\footnote{http://www.speech.cs.cmu.edu/cgi-bin/cmudict}  as a good choice for the lexicon including three levels of stress. The input phoneme sequence has 70 dimensions. The encoder contains three convolutional layers with filter width equals 5 and 256 channels. The HTK\footnote{http://htk.eng.cam.ac.uk/}  aligns the phoneme sequence with the speech samples using forced-alignment. We have replaced the monophone with the triphone sequence but not that much change in results. In addition, we have tried summation of the activations of each layer in encoder as the latent feature but the results were worse.

The dilations of each generator are $1, 2, 4, 8$, and $16$. The channel size for dilation, residual, and skip-connection were set to $256$. Adam optimizer [27] is used for training with the learning rate initiating from $10^{-3}$ and decaying every $50k$ iteration by a factor of $0.5$.

\subsection{Evaluation metrics}
\label{exp-eval-metr}
The evaluation metrics are signal-to-noise ratio (SNR) in time domain and logarithmic spectral distortion (SD) which are defined as follows:

\begin{equation}  
\scriptstyle
{SNR}_{[dB]}=10{log}_{10}\frac{\sum^T_{t=1}{{s(t)}^2}}{\left|\sum^T_{t=1}{{s(t)}^2}-\sum^T_{t=1}{{\hat{s}(t)}^2}\right|}\  
\end{equation} 

\begin{equation} 
\scriptstyle
{SD}_{[dB]}=\frac{1}{N}\sum^N_{n=1}{\sqrt{\frac{1}{F}\sum^F_{f=1}{{\left[20{log}_{10}\frac{\left|S(f,n)\right|}{\left|\hat{S}(f,n)\right|}\right]}^2}}}\ ,  
\end{equation} 

where $S(f,n)$ and $\hat{S}(f,n)$ are spectrograms of the target signal and the generated signal, respectively. The spectrograms are calculated by 16 ms frame length, 1 ms shift and Hanning window. In addition, because the human auditory perception is based on the Mel spectrogram representation, we considered Mel spectral distortion (MSD) as the third quantitative metric for the objective evaluation. The MSD is calculated similar to the SD, replacing the linear spectrogram with the 40-filters Mel spectrogram, which is obtained by 25 ms window length, and 5 ms shift. In addition, we calculated the SNR in the linear spectrogram domain. We did not mention the spectrogram SNR results because with two digits precision they are the same as the time domain ones. The generation is time consuming in the proposed model because the speech is synthesized sample-by-sample and sequentially. Therefore, in addition to the above-mentioned metrics, the training and the synthesis time are discussed, too. 

\subsection{Results and discussions}
\label{exp-results}

\subsubsection{Objective evaluation}
\label{objective-eval}

First experiment investigated the effect of the wavelet analysis/synthesis on the quality of speech without engaging any neural network model. The average results on 100 test set utterances with 95\% confidence interval are reported in the first row of Table \ref{eval-snr-table} as the extreme case for evaluations. For SNR, higher value shows more accurate model; whereas for both SD and MSD, lower value means better performance. As shown in Table \ref{eval-snr-table}, the subband decomposition/reconstruction results provides near perfect performance.

Moreover, we compared the subband with the fullband speech synthesizer. The fullband term means that the model prediction $\hat{s}$ is the speech signal in its full frequency range. Hence, there is no subband decomposition. Therefore, one complex signal generator models the probability distribution. Basically, the two models are exactly the same, except in the fullband TTS, $K=24$ dilation layers are defined with 4 stacks of 1, 2, 4, …, 32 dilations in our experiments; while in the subband TTS, the number of dilation layers is much lower than the fullband ($K=5$). Fast WaveNet algorithm [17] is utilized for the synthesis of both models.

We compare the two models by conditioning on the phoneme sequence as the conditional features in two cases: \emph{teacher-forcing} and \emph{synthesis}. The teacher-forcing means feeding the previous clean subband samples as $x$ to the model and evaluating the accuracy of the prediction of the next sample. In fact, as an input of the generator, $x$ refers to the previous clean subband samples $s_l$ and previously generated samples $\hat{s}_l$ in teacher-forcing and synthesis, respectively. As depicted in Table \ref{eval-snr-table}, the subband model performs significantly better than the fullband one in teacher-forcing. For synthesis, the results are somehow comparable. In fact, the results of synthesis are not satisfying for both subband and fullband models, which is probably due to the lack of acoustic conditional features.

\begin{table}
  \caption{Evaluation results (mean $\mathrm{\pm}$ 95\% CI) for 100 test set utterances}
  \label{eval-snr-table}
  \centering
\footnotesize
    \begin{tabular}{p{0.5cm}p{1cm} p{1.4cm} p{1.4cm} p{1.4cm}}
    \toprule
             &  & SNR [dB] & SD [dB] & MSD [dB] \\ 
    \midrule 
    \multicolumn{2}{p{1.9cm}}{wavelet dec/rec} & 41.5 $\mathrm{\pm}$ 1.14 & 0.61 $\mathrm{\pm}$ 0.01 & 0.08 $\mathrm{\pm}$.002 \\ \hline 
     \multirow{2}{*}{t.f.} & Subband & \textbf{23.5 $\boldsymbol{\mathrm{\pm}}$ 0.31} & \textbf{4.3 $\boldsymbol{\mathrm{\pm}}$ 0.02} & \textbf{2.5 $\boldsymbol{\mathrm{\pm}}$ 0.01} \\  
\textit{} & Fullband & 18.8 $\mathrm{\pm}$ 0.47 & 8.1 $\mathrm{\pm}$ 0.03 & 5.5 $\mathrm{\pm}$ 0.04 \\ \hline 
      \multirow{2}{*}{syn.} & Subband & 4.0 $\mathrm{\pm}$ 0.88 & \textbf{13.3 $\boldsymbol{\mathrm{\pm}}$ 0.01} & \textbf{10.0 $\boldsymbol{\mathrm{\pm}}$ 0.10} \\ 
\textit{} & Fullband & \textbf{5.2 $\boldsymbol{\mathrm{\pm}}$ 0.93} & 15.2 $\mathrm{\pm}$ 0.10\textbf{} & 11.8 $\mathrm{\pm}$ 0.11 \\ 
      \bottomrule 
      \end{tabular}
\end{table}

\subsubsection{Generated samples}

For subjective comparison, Mel spectrograms of a randomly selected sample together with the the teacher-forcing outputs of the fullband and subband models are plotted in Figure 3. The spectrograms are compatible with the objective evaluation explained in subsection \ref{objective-eval}. Figure 3 shows the near perfect teacher-forcing results. Furthermore, the subband teacher-forcing outputs confirm the potential of the proposed model and the priority of the subband over the fullband one. The corresponding wave files are also available on our Github page (\url{https://github.com/AzamRabiee/subband-TTS}).

\begin{figure}
  \centering
    \includegraphics[width=0.45\textwidth]{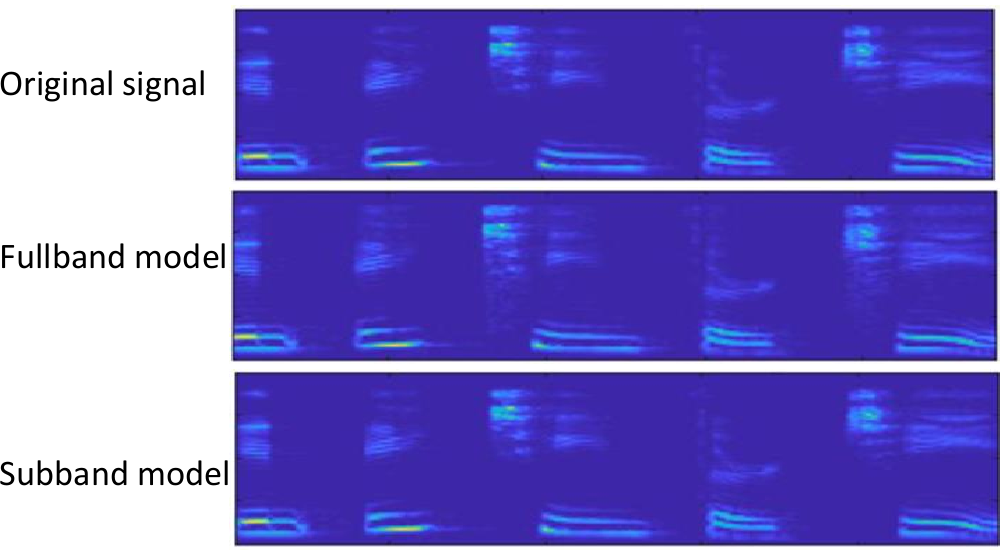}
    \caption{Mel spectrograms of generated signals by fullband and subband models with teacher-forcing. The sample is selected randomly.}
\end{figure}

\subsubsection{Synthesis and train time}

The generation rate of the subband TTS is measured in two cases: \emph{sequential} and \emph{parallel}. For the earlier experiment, the speech signal is decomposed into subband signals; and they are kept in the original sampling rate, which is 16 kHz in our case. Thus, the samples are \emph{redundant}. Obviously, without parallelization, the generation rate of the redundant samples should be 8 times less than the fullband because there are 8 subband signals in the experiments. Nevertheless, since the complexity of each subband network is less than the fullband one, it is 2.5 times slower for an Intel Xeon CPU with 32 cores. For the parallel experiment, in addition to the parallel subband generation, the subband signals are downsampled by a factor of 2 as it is allowed by the first level of the wavelet transform. Hence, the subband $s_1$ is not redundant any more. The generation rate of the parallel subband implementation is 150Hz (higher than 110Hz for the fullband model). Even though the parallelization and the downsampling speed up the synthesis, but it is still not that much far from the fullband model and much lower than the non-autoregressive models [18]-[20]. Still the divide and conquer idea presented in this paper is applicable on non-autoregressive networks. 

Both subband and fullband models need less than a day (around 18 hours) for training up to an admissible output quality (200k global steps in our experiments) on a Titan X (Pascal) GPU with 12 GB memory. Such a fast training is because of their fully CNN architecture, which is much better than the RNN-based TTS, e.g. Tacotron [6]. It is reported that an implementation of the Tacotron takes 12 days (877K iterations) on a GTX 1080 Ti\footnote{https://github.com/keithito/tacotron}. Note that the number of iterations is still much less than the original Tacotron reported by Google (2M iterations) [6].

\section{Conclusion}
\label{conclusion}

We proposed a subband time-domain TTS system inspiring from the WaveNet. The main differences of our TTS with the WaveNet are twofold: first, rather than a complex deep neural network for modeling the probability distribution of the speech signal, we designed separate (but integrated) networks for each subband signal, which has much simple architecture per subband and could estimate the probability distributions of the subband signals accurately. Second, WaveNet vocoder uses pretrained acoustic feature prediction, and the original WaveNet paper uses linguistic features plus $log(f_0)$ predicted by its text-analysis and $f_0$ pretrained front-end modules; while an encoder in our system extracts the latent linguistic features from the phoneme sequence input in a nearly end-to-end way, which is more preferred. The force alignment should be replaced by an attention mechanism for automatic aligning to have a fully end-to-end model. Still enriching the conditional features by acoustic features beside the current linguistic features is unavoidable. 

\section{Acknowledgements}

This work was supported by Institute of Information \& Communications Technology Planning \& Evaluation (IITP) grant funded by the Korea government(MSIT) [2016-0-00562(R0124-16-0002), Emotional Intelligence Technology to Infer Human Emotion and Carry on Dialogue Accordingly]



\section{References}
{\small 
\hangindent=0.5cm [1]	A. J. Hunt and A. W. Black, "Unit selection in a concatenative speech synthesis system using a large speech database," in Proc. IEEE International Conference on Acoustics, Speech and Signal Processing (ICASSP), 1996, vol. 1, pp. 373–376.

\noindent \hangindent=0.5cm [2]	N. S. Kim and S. S. Park, "Discriminative training for concatenative speech synthesis," IEEE Signal Process. Lett., vol. 11, no. 1, pp. 40–43, 2004.

\noindent \hangindent=0.5cm [3]	H. Zen, K. Tokuda, and A. W. Black, "Statistical parametric speech synthesis," Speech Commun., vol. 51, no. 11, pp. 1039–1064, 2009.

\noindent \hangindent=0.5cm [4]	Y.-J. Hu and Z.-H. Ling, "DBN-based spectral feature representation for statistical parametric speech synthesis," IEEE Signal Process. Lett., vol. 23, no. 3, pp. 321–325, 2016.

\noindent \hangindent=0.5cm [5]	Z.-C. Liu, Z.-H. Ling, and L.-R. Dai, "Statistical Parametric Speech Synthesis Using Generalized Distillation Framework," IEEE Signal Process. Lett., vol. 25, no. 5, pp. 695–699, 2018.

\noindent \hangindent=0.5cm [6]	Y. Wang et al., "Tacotron: A fully end-to-end text-to-speech synthesis model," in Proc. Interspeech, 2017, pp. 4006–4010.

\noindent \hangindent=0.5cm [7]	H. Tachibana, K. Uenoyama, and S. Aihara, "Efficiently trainable text-to-speech system based on deep convolutional networks with guided attention," in Proc. IEEE International Conference on Acoustics, Speech and Signal Processing (ICASSP), 2018, pp. 4784–4788.

\noindent \hangindent=0.5cm [8]	A. Van Den Oord et al., "Wavenet: A generative model for raw audio," CoRR, vol. abs/1609.0, 2016.

\noindent \hangindent=0.5cm [9]	G. Lai, B. Li, G. Zheng, and Y. Yang, "Stochastic WaveNet: A Generative Latent Variable Model for Sequential Data," arXiv Prepr. arXiv1806.06116, 2018.

\noindent \hangindent=0.6cm [10]	K. Qian, Y. Zhang, S. Chang, X. Yang, D. Florêncio, and M. Hasegawa-Johnson, "Speech enhancement using Bayesian wavenet," in Proc. Interspeech, 2017, pp. 2013–2017.

\noindent \hangindent=0.6cm [11]	D. Rethage, J. Pons, and X. Serra, "A Wavenet for speech denoising," in Proc. IEEE International Conference on Acoustics, Speech and Signal Processing (ICASSP), 2018, pp. 5069–5073.

\noindent \hangindent=0.6cm [12]	A. Tamamori, T. Hayashi, K. Kobayashi, K. Takeda, and T. Toda, "Speaker-dependent WaveNet vocoder," in Proc. Interspeech, 2017, vol. 2017–Augus, pp. 1118–1122.

\noindent \hangindent=0.6cm [13]	T. Hayashi, A. Tamamori, K. Kobayashi, K. Takeda, and T. Toda, "An investigation of multi-speaker training for wavenet vocoder," in Proc. Automatic Speech Recognition and Understanding Workshop (ASRU), 2017, pp. 712–718.

\noindent \hangindent=0.6cm [14]	W. Ping et al., "Deep voice 3: Scaling text-to-speech with convolutional sequence learning," in Proc. 6th International Conference on Learning Representations (ICLR), 2018, vol. 79, no. 14, pp. 1094–1099.

\noindent \hangindent=0.6cm [15]	T. Yoshimura, K. Hashimoto, K. Oura, Y. Nankaku, and K. Tokuda, "Mel-cepstrum-based quantization noise shaping applied to neural-network-based speech waveform synthesis," IEEE/ACM Trans. Audio, Speech, Lang. Process., vol. 26, no. 7, pp. 1173–1180, 2018.

\noindent \hangindent=0.6cm [16]	J. Shen et al., "Natural TTS synthesis by conditioning WaveNet on Mel spectrogram predictions," in Proc. IEEE International Conference on Acoustics, Speech and Signal Processing (ICASSP), 2018, pp. 4779–4783.

\noindent \hangindent=0.6cm [17]	T. Le Paine et al., "Fast wavenet generation algorithm," CoRR, vol. abs/1611.0, 2016.

\noindent \hangindent=0.6cm [18]	A. Van Den Oord et al., "Parallel WaveNet: Fast high-fidelity speech synthesis," CoRR, vol. abs/1711.1, 2017.

\noindent \hangindent=0.6cm [19]	W. Ping, K. Peng, and J. Chen, "Clarinet: Parallel wave generation in end-to-end text-to-speech," in ICLR 2019 Conference, 2019.

\noindent \hangindent=0.6cm [20]	R. Prenger, R. Valle, and B. Catanzaro, "WaveGlow: A Flow-based Generative Network for Speech Synthesis," in IEEE International Conference on Acoustics, Speech and Signal Processing (ICASSP), 2019.

\noindent \hangindent=0.6cm [21]	T. Inoue, S. Hara, and M. Abe, "A hybrid text-to-speech based on sub-band approach," in Proc. Signal and Information Processing Association Annual Summit and Conference (APSIPA), 2014 Asia-Pacific, 2014, pp. 1–4.

\noindent \hangindent=0.6cm [22]	T. Okamoto, K. Tachibana, T. Toda, Y. Shiga, and H. Kawai, "An investigation of subband WaveNet vocoder covering entire audible frequency range with limited acoustic features," in Proc. IEEE International Conference on Acoustics, Speech and Signal Processing (ICASSP), 2018, pp. 5654–5658.

\noindent \hangindent=0.6cm [23]	Y. Lee, A. Rabiee, and S.-Y. Lee, "Emotional End-to-End Neural Speech Synthesizer," in Workshop Machine Learning for Audio Signal Processing at NIPS (ML4Audio@NIPS17), 2017.

\noindent \hangindent=0.6cm [24]	I. Daubechies, Ten lectures on wavelets, vol. 61. Siam, 1992.

\noindent \hangindent=0.6cm [25]	M. Farge, "Wavelet transforms and their applications to turbulence," Annu. Rev. Fluid Mech., vol. 24, no. 1, pp. 395–458, 1992.

\noindent \hangindent=0.6cm [26]	Y. Wang et al., "Style tokens: Unsupervised style modeling, control and transfer in end-to-end speech synthesis," in Proc. International Conference on Machine Learning (ICML), 2018.

\noindent \hangindent=0.6cm [27]	D. P. Kingma and J. L. Ba, "Adam: Amethod for stochastic optimization," in Proc. of the 3rd International Conference on Learning Representations (ICLR), 2015.

}


\end{document}